\documentstyle[12pt]{article}
\newcommand{\order}[1]{\mbox{${\cal O}(#1)$}}
\newcommand{\half}{{1 \over 2}}
\newcommand{\third}{{1 \over 3}}
\newcommand{\threehalf}{{3 \over 2}}
\newcommand{\fivehalf}{{5 \over 2}}
\newcommand{\sevenhalf}{{7 \over 2}}

\newcommand{\equa}[1]{(\ref{#1})}

\begin{document}
{\LARGE\bf
\begin{center}
Higher Resonance Contributions\\
to the Adler--Weisberger Sum Rule\\
in the Large $N_c$ Limit
\end{center}}

{\large
\vskip 1cm
\begin{center}
A. Wirzba$^1$, M. Kirchbach$^1$ and D.O. Riska $^2$\\[4mm]
$^1$ Institut f\"ur Kernphysik, Technische Hochschule Darmstadt,\\
D--64289 Darmstadt, Germany\\[2mm]
$^2$ Department of Physics, SF--00014 University of Helsinki, Finland
\end{center}}

\vskip 3cm
\centerline{\bf Abstract}

\noindent
We determine the $N_c$--dependence of the resonance contributions to
the Adler--Weisberger sum rule for the inverse square $1/g_A^2$ of the
axial charge coupling constant and show that in the large $N_c$ limit
the contributions of the Roper-like excitations scale as
$\order{1/N_c}$.  Consistency with the $1/N_c^2$ scaling of the
$1/g_A^2$ term in the sum rule requires these contributions to cancel
against each other.

\vfill
\noindent hep-ph/9311299

\newpage
\section{Introduction}
\setcounter{equation}{0}

If the the axial charge commutator in the Adler-Weisberger relation
for the axial vector coupling constant is saturated by a discrete set
of pion-nucleon resonances (with spin $j$, parity $P$ and isospin $I$)
the sum rule takes the following form
\cite{Della 67},
\cite{Maz 87}
\begin{equation}
  \frac{1}{g_A^2}    = 1 + \sum_{(j^P,I)}
             f_I\,
            { { g_{j I}^{(\pm)}}^2 \over g_{\pi NN}^2 }\, m_N^2\,
              \frac{(m_{j I}^2- m_N^2)^{2j-3}}{(m_N m_{j I})^{2j-1}}\,
               \frac{(m_{j I} + \eta \,  m_N)^2}{2^{j-\sevenhalf}}\,
               \frac{[(j+\half)!]^2}{(2j+1)!}\ .
    \label{sumrule}
\end{equation}
In this expression $g_{\pi NN}$ stands for the pseudoscalar pion
nucleon coupling constant, $g_{j I}^{(\pm)}$ is the (dimensionless)
coupling constant for the spin--$j$ (isospin--$I$) resonance coupling
to the pion-nucleon system in the relativistic Rarita--Schwinger
representation \cite{DeAlf 73}, and $m_N$ and $m_{j I}$ denote the the
nucleon and resonance masses, respectively. Subsequently, $f_\pi$
denotes the pion weak decay coupling constant.

The quantities $g_A$, $m_N$ and $m_{j I}$ scale with the number of
color degrees of freedom~\cite{tHooft 74,Witten 79} as $\order{N_c}$,
whereas $f_\pi\sim \order{\sqrt{N_c}}$ and $g_{\pi NN} \sim
\order{N_c^\threehalf}$.  The mass differences $m_{j I}
- m_N$ are either of order $\order{N_c^0}$ or of order
$\order{1/N_c}$. The former scaling behavior is the usual situation
for resonances in meson-baryon scattering~\cite{Witten 79} and applies
e.g.\ for the Roper-like resonances as well as for the odd-parity
states. The latter scaling behavior is special for e.g.\ the
$\Delta(1232)$ and its partners in the $(j^{+},I(j))$ tower that
belongs to the contracted SU(4)-symmetry, which applies for the baryon
states in strong coupling models~\cite{Bar 84,Ger 84} and the $1/N_c$
expansion~\cite{Dashen 93,Jen 93}.  The isospin factor $f_I$ takes the
value $1$ for the $I=\half$ resonances and the value $-\third$ for the
$I=\threehalf$ ones. (Note that resonances with $I>\threehalf$ cannot
couple to the pion-nucleon system).  The parameter $\eta $ in
\equa{sumrule} takes the values $\pm$ and corresponds to the $\pm$
structure in the exponent of the parity $P= (-1)^{j\pm 1/2}$ of the
intermediate resonance.  For example, the sign ``+" corresponds to the
odd-parity resonances (with $(-1)^{j+\half}=-1$) as well as for the
$\Delta$--resonance and its Roper like excitation, whereas the ``$-$"
sign has to be used for the Nucleon-Roper, the Delta-odd-parity
resonant excitations etc.

With the exception of the coupling constants $g_{j I}^{(\pm)}$ the
scaling with $N_c$ of all the quantities that appear in
eq.\equa{sumrule} is known. The left hand side of eq.\equa{sumrule} is
of order $\order{1/N_c^2}$. The right hand side, however, because of
its first summand, has an apparent order $\order{N_c^0}$ term. The
resonance contributions on the right hand side of eq.\equa{sumrule}
therefore have to combine so as to (a) exactly cancel the first
summand (1) and to (b) give the required $\order{1/N_c^2}$ behavior.
This requirement was used in ref.\cite{Kak 81}
to get the $N_c$ scaling of the
coupling constants $g_{j I}^{(\pm)}$. We here
show that the quark model implies that the coupling constants for
the Roper-like resonances scale as $N_c$ and not as $N_c^{3/2}$
as inferred from the sum rule in ref.\cite{Kak 81}.
This $N_c$ dependence implies that these resonances
give contributions of order $1/N_c$ to the sum rule, and thus
consistency with the r.h.s.\ requires these contributions to cancel against
each other. That cancellation is made possible by the opposite
signs of the contributions from the Roper
and $\Delta$-Roper resonances
in the sum rule.

\section{$N_c$ scaling of the coupling constants}

In the relativistic Rarita-Schwinger representation $\pi N$-resonance
vertices are~\cite{DeAlf 73}:
\begin{eqnarray}
    \langle N(p) | \chi^\beta | j(\mbox{$p$+$q$})\rangle
     &=& \frac{g_{j I}^{(+)}}{m_N^{j-\half}}
      \bar u(p) u_{\mu_1\cdots \mu_{j-\half}}\!\mbox{($p$+$q$)}\
      q'^{\mu_1}\cdots q'^{\mu_{j-\half}}\, \xi^{+}_{\half}
     \left\{\begin{array}{r}\tau^\beta \xi_{\half}\\
                         \ \xi_{\threehalf}^\beta\end{array}\right\}  ,
 \label{vertexp}\\
    \langle N(p) | \chi^\beta | j(\mbox{$p$+$q$})\rangle
     &=& \frac{ig_{j I}^{(-)}}{m_N^{j-\half}}
      \bar u(p)\gamma_5 u_{\mu_1\cdots
      \mu_{j-\half}}\!\mbox{($p$+$q$)}\
      q'^{\mu_1}\cdots q'^{\mu_{j-\half}}\, \xi^{+}_{\half}
     \left\{\begin{array}{r}\tau^\beta \xi_{\half}\\
                    \ \xi_{\threehalf}^\beta\end{array}\right\}  ,
 \label{vertexm}
\end{eqnarray}
where $u(p)$ and $u_{\mu_1\cdots \mu_{j-\half}}\!\mbox{($p$+$q$)}$
are in turn
the spinor and and totally symmetric
tensor-spinor wave functions that describe the proton
and the spin--$j$ resonance, $\xi_{\half}$,
$\xi_{\threehalf}^\beta$ are the wave functions of the isospin--$\half$
or $\threehalf$ resonances, respectively, and $\xi_{\half}^{+}$ is the
isospin--$\half$ wave function of the proton.  The two equations
correspond to resonances of parity $(-1)^{j+\half}$ and
$(-)^{j-\half}$, respectively. The four-momentum $q'_\mu$ is defined
as
\begin{equation}
       q'_\mu = q_\mu - \frac{q(p+q)}{m_{jI}^2} (p+q)_\mu \ .
\end{equation}

Compare now the $N_c$ behavior of the non-relativistic reduction (in
the baryon degrees of freedom) of eqs.\equa{vertexp} and
\equa{vertexm} with the results for the non-relativistic quark model
in the large $N_c$ limit.  In this limit the non-relativistic
reduction is well justified, since the baryon masses scale as
$\order{N_c}$ and can be made arbitrarily heavy. The naive quark model
can also be relied upon in this limit as we only need the leading
$N_c$ behavior of the resonance couplings, which is
model-independent~\cite{Dashen 93}.  In the non-relativistic limit the
{\em spin-isospin}-structure of the one-body vertex operators sandwiched
between the proton state $|N\rangle $ and a (by the transition
allowed) baryon state $|B_j \rangle$ can take only one of the
following four forms: a scalar-isoscalar sum
\begin{equation}
   \langle N| \frac{G}{f_\pi}\sum_{\nu=1}^{N_c} {\bf 1_\nu}
   |B_j\rangle
      \sim
       \left\{ \begin{array}{l} \order{\sqrt{N_c}} \\
                               \order{N_c^0} \end{array} \right . \ ,
 \label{sis}
\end{equation} a
pseudoscalar-isovector sum
\begin{equation}
     \langle N |\frac{G}{f_\pi}\sum_{\nu=1}^{N_c}
     \vec\sigma_\nu \cdot \vec q\,\tau^\beta_\nu |B_j \rangle
    \sim
       \left\{\begin{array}{l} \order{\sqrt{N_c}} \\
                               \order{N_c^0} \end{array} \right . \ ,
 \label{psiv}
\end{equation}
a scalar-isovector sum
\begin{equation}
  \langle N |\frac{G}{f_\pi}\sum_{\nu=1}^{N_c} \tau^\beta_\nu
  |B_j\rangle
  \sim
        \left\{\begin{array}{l} \order{1/\sqrt{N_c}} \\
                               \order{1/{N_c}} \end{array} \right . \ ,
 \label{siv}
\end{equation}
or
a pseudoscalar-isoscalar sum
\begin{equation}
  \langle N |
   \frac{G}{f_\pi}\sum_{\nu=1}^{N_c} \vec\sigma_\nu\cdot \vec q \,
   |B_j\rangle \sim
       \left\{\begin{array}{l} \order{1/\sqrt{N_c}} \\
                               \order{1/ {N_c}} \end{array} \right . \ .
 \label{psis}
\end{equation}
Here the index $\nu$ counts the $N_c$ different quarks of the proton
and the spin--$j$ resonance.  In \equa{psis} $G$ is a pion-quark
coupling strength which scales as $\order{N_c^0}$.  The pion decay
constant $f_\pi$ appears through the normalization of the pion field
in the vertex operator.  The spatial vector $\vec q$ is the pion
three-momentum in the resonance rest frame which also scales as
$\order{N_c^0}$ for fixed kinematics. In the equations above the upper
scaling behavior on the r.h.s.\ applies to the case when $B_j$ is the
nucleon, or a state in which all quarks are in their spatial ground
state (e.g.\ the lowest $\Delta$ resonance). The lower scaling behavior
applies for such baryon states with one or two quarks  in an
excited spatial state (e.g.\ the Nucleon- or $\Delta$-Roper ).  In that
case the symmetrization of the color-singlet baryon state introduces
an additional normalization factor $1/\sqrt{N_c}$~\cite{Col 79}, which
leads to a similar extra factor in the $N_c$ dependence of the matrix
element.  This is consistent with the fact that -- because of
unitarity -- the $\pi N \to \pi N$ scattering amplitude can at most
scale as $\order{N_c^0}$: The direct Born diagram with a nucleon as
intermediate state scales by itself as $\order{N_c}$ since both
vertices scale as $\order{\sqrt{N_c}}$. In ref.\cite{Dashen 93} it was shown
that this leading $N_c$ scaling cancels when the crossed Born diagram
{\em and} the corresponding direct and crossed diagrams with the (in the
large $N_c$ limit degenerate $SU(4)$ partner) $\Delta$ are
included. However, in case the intermediate state in the $\pi N \to
\pi N$ scattering is a (with the nucleon non-degenerate)
baryon-resonance, the direct Born diagram cannot be cancelled any
longer in this way. Thus, in order to be consistent with unitarity,
the vertex itself can at most scale as $\order{N_c^0}$
in agreement with the scaling
behavior derived from the normalization of the excited baryon
state~\cite{Col 79}.

Note that in the vertex sum \equa{psiv} the single quark
$\vec \sigma_\nu \cdot \vec q \tau_\nu^\beta$ contributions add up
coherently and lead to the the same $N_c$ scaling as in eq.\equa{sis},
whereas the other two vertex sums
\equa{siv} and \equa{psis} are suppressed by $1/N_c$ since the
quark contributions only add up destructively. In fact only the vertex
sums \equa{psiv} and \equa{siv} contribute here because of the isovector nature
of the pion coupling. The $N_c$
behavior of the pion-proton vertices in the naive quark model
is thus determined.

The non-relativistic reduction of the
Rarita-Schwinger-type vertex matrix elements
\equa{vertexp} and \equa{vertexm} is standard.
Using the {\em totally symmetric} nature of the $j$-$\half$ tensor
spinors one can identify -- in the non-relativistic limit and in spin-isospin
space --
the matrix elements \equa{vertexp} with the pseudoscalar-isovector
structure, eq.\equa{psiv}, of the naive
quark model, if $j$-$\half$ is odd, and with the scalar-isovector one,
eq.\equa{siv}, if $j$-$\half$ is even.  For the vertex matrix
element \equa{vertexm} the fact has to be taken into account that the
$\gamma_5$ Dirac matrix links one upper and one lower component of the
proton and the spin--$j$ resonance and therefore induces an extra
factor $\vec\sigma\cdot (\vec p+\vec q\,)/2 m_{jI} -\vec\sigma\cdot
\vec p /2 m_{N} = \vec \sigma\cdot \vec q/2m_N +
\order{1/N_c^2}$~\footnote{Note
that in the rest frame of the resonance $\vec q = {\vec q}\,'$.}.
Therefore the vertex matrix element \equa{vertexm}  leads to the
pseudoscalar-isovector structure, eq.\equa{psiv}, for $j$-$\half$ even
and to the scalar-isovector one, eq.\equa{siv}, for $j$-$\half$
odd. Thus, the $N_c$ behavior of the $g_{j I}^{(\pm)}$'s can now be
predicted.

The scaling of the Roper resonance coupling
$g_{\half \half}^{(-)}$~\footnote{The ``$-$'' sign in the notation of
ref.\cite{DeAlf 73} is reflecting the ``$-$'' sign in $j-\half$ and
should not be mixed up with the parity of the Roper state which of
course is positive.} follows via eqs.\equa{vertexm} and \equa{psiv} from
\begin{equation}
      \frac{g_{\half \half}^{(-)}}{ 2 m_N} \sim \frac{G}{f_\pi\sqrt{N_c}}
\times {N_c}
           \sim \order{1}
\end{equation}
and is thus $\order{N_c}$.
The scaling of the odd-parity-nucleon-like coupling
$g_{\half \half}^{(+)}$  results
from
\begin{equation}
     g_{\half \half}^{(+)} \sim \frac{G}{f_\pi\sqrt{N_c}}
     \sim \order{1/{N_c}}
\end{equation}
(see eqs.\equa{vertexp} and \equa{siv}) and is $\order{1/{N_c}}$.
The scaling behavior of the $\Delta$ (ground state) coupling
$g_{\threehalf \threehalf}^{(+)}$ can be deduced from
\begin{equation}
  \frac{g_{\threehalf \threehalf}^{(+)}}{m_N} \sim \frac{G}{f_\pi}N_c
\sim \order{\sqrt{N_c}}
\end{equation}
as $N_c^\threehalf$, whereas the coupling $g_{\threehalf'
\threehalf}^{(+)}$ of the Roper-like $\Delta$ scales -- because of the
$1/\sqrt{N_c}$ normalization -- only as $N_c$:
\begin{equation}
  \frac{g_{\threehalf' \threehalf}^{(+)}}{m_N} \sim \frac{G}{f_\pi\sqrt{N_c}}
 \times {N_c}
\sim \order{N_c^0}
\end{equation}
(see eqs.\equa{vertexp} and \equa{psiv}).  The coupling constant
$g_{\threehalf \threehalf}^{(-)}$ of the odd-parity
$\Delta$--resonance also behaves as $N_c$ (see eqs.\equa{vertexm} and
\equa{siv}), since
\begin{equation}
       \frac{g_{\threehalf \threehalf}^{(-)}}{m_N^2}
         \sim \frac{G}{f_\pi\sqrt{N_c}}
          \sim \order{1/{N_c}}\ .
\end{equation}
In general, comparing the non-relativistic reduction of
eqs.\equa{vertexp} or \equa{vertexm} with the $N_c$-counting of
eqs.\equa{psiv} or \equa{siv} one can derive the following relations
for the coupling constants of the $(j^{\pm},I)$ states with
$I$=$\half$ or $\threehalf$:~\footnote{With the exception of the lowest
$\Delta$ state for which the coupling constant behave as
$\order{N_c^\threehalf}$ and not as $\order{N_c}$.}
\begin{equation}
 \begin{array}{lclcl}
  g_{j I}^{(+)}
        &=&  \order{N_c^{j-\half}} &{\rm for} &   P=(-1)^{j+\half}=+ \\
  g_{j I}^{(+)}
        &=&  \order{N_c^{j-\threehalf} }  &{\rm for} & P=(-1)^{j+\half}=- \\
  g_{j I}^{(-)}
        &=&  \order{N_c^{j+\half} }  &{\rm for} & P=(-1)^{j-\half}=+ \\
  g_{j I}^{(-)}
        &=&  \order{N_c^{j-\half} }  &{\rm for} & P=(-1)^{j-\half}=- \ ,
 \end{array}
\end{equation}
(see the 5th row in Table~1).

\section{$N_c$ scaling of the resonance contributions to the sum rule}

The $N_c$ scaling of all the quantities that enter on the right hand
side of the Adler-Weisberger sum rule \equa{sumrule} is now
determined. In the 7th row of Table~1 the contributions of the single
resonances with $j\leq 7/2$ are listed.  Using the results of Table 1
one finds  that (a) the $\Delta(1232)$ contributes
to leading order $\order{N_c^0}$
to the r.h.s.\ of the sum rule, (b) the $N$-Roper, the $\Delta$-Roper,
the $(\threehalf^{+},\half)$, the
$(\fivehalf^{+},\half)$, the $(\half^{+},\threehalf)$ etc.\
contribute to  next-to-leading order, $\order{1/N_c}$, and (c)
the contribution of e.g.\ the odd-parity $(\half^{-},\half)$ nucleon
excitation is suppressed by $\order{1/N_c^3}$.

Since the contributions of the the radial excitations of the nucleon-
and the $\Delta$-resonances contribute to next-to-leading order,
$\order{1/N_c}$, they have to cancel in order to keep consistency with
the scaling of the l.h.s.\ of the sum rule. The contributions from
these states therefore have to be considered with some care when the
Adler-Weisberger sum-rule is applied to large $N_c$ models as the
Skyrme-model (or its extensions) or large $N_c$ quark models.
The
Roper-like states, on the one side, and the $\Delta(1232)$ resonance,
on the other side, contribute for rather different reasons significantly to
the r.h.s.\ of
the sum rule in eq.\equa{sumrule}. In the large $N_c$ limit the latter is
degenerate with the nucleon (the mass splitting goes as
$\order{1/N_c}$) and has a width that vanishes with $1/N_c^2$. The
Ropers, on the other hand, do not become degenerate with the nucleon
in the large $N_c$ limit as the splitting is $\order{N_c^0}$, but
their widths are also of order $\order{N_c^0}$ and therefore large
enough to leave strength left at the nucleon pole. The masses of the
other resonances are order $\order{N_c^0}$ above the nucleon pole and
have widths that are only of order $\order{1/N_c^2}$. Therefore they
have no strength at the nucleon pole in the large $N_c$ limit and may
be neglected.

The widths of the spin--$j$ isospin--$I$
resonances are given by the expression~\cite{Della 67,DeAlf 73}:
\begin{equation}
 \Gamma_{j I}^{(\pm)}= 3 |f_I| \frac{{g_{j I}^{(\pm)}}^2}
        {4\pi m_N^{2j-1}}\,
  \frac{2^{j-\half} [(j-\half)!]^2}{(2j)!}
 \frac{E_N + \eta \, m_N}{m_{j I}} ({\vec p\,}^2)^j\ .
 \label{widths}
\end{equation}
Here $\vec p$ and $E_N$ are the three-momentum and the energy of the
nucleon in the $j$-resonance rest frame, respectively.  For fixed
kinematics $E_N$ scales as $\order{N_c}$, whereas $|\vec p|$ normally
scales as $\order{N_c^0}$. Note, however, that the nucleon kinetic
energy $E_N-m_N$ scales as $\order{1/N_c}$, which is obvious in the
non-relativistic limit. Furthermore for the $\Delta(1232)$ (and its
partners in the $(j^{+},j)$ tower) the three-momentum $\vec p$ does
not scale as $\order{N_c^0}$, but rather as $1/N_c$.  From the
expression \equa{widths}, it then follows that the widths of all the
Roper like excitations scale as $\order{N^0_c}$, whereas those of the
odd-parity resonances scale as $\order{1/N^2_c}$ (see Table~1).
Finally, as pointed out above, the width of the $\Delta(1232)$ falls
with $N_c$ at least as $\order{1/N_c^2}$.

\section{Conclusions}

The contribution $C_{\Delta }$ of the $\Delta$ resonance
to the right hand side of the Adler--Weisberger sum rule~\equa{sumrule}
is:
\begin{equation}
C_{\Delta}  =  - {2\over 9}
          \frac{{g_{\threehalf \threehalf }^{(+)}}^2}
                                 {g_{\pi NN}^2}
{{(m_{\Delta} + m_N)^2}\over m_{\Delta}^2}\, .
\end{equation}
In the large $N_c$ limit when $m_{\Delta}\to m_N$ one finds that
$C_{\Delta} \to -1$ \cite{Maz 87,Ya 92} in accordance with the
prediction of the contracted SU(4) spin--flavor symmetry.
The contribution $C_\Delta$ alone
is therefore sufficient to cancel
the terms of order $N_c^0$ in the sum rule
eq.\equa{sumrule} exactly.  However, as shown in
section~3,
the contributions from the
Roper--like, as well as the $D_{13}$ negative parity resonances in the
large $N_c$ limit
scale as $1/N_c$ and consistency therefore requires these to
cancel against each other~\footnote{The importance of the Nucleon-Roper
resonance for the saturation of the chiral axial charge commutator has
already been emphasized in \cite{Ki 91}.}.
This behavior of the Roper-like states is in line with the unitarity
limit, $\order{N_c^0}$, of the total cross sections
$\sigma_{\rm tot}^{\pi^{\pm}p}$ in $\pi^{\pm}$-proton scattering which
appear on the r.h.s.\ of the integral form of
the Adler-Weisberger relation
\begin{equation}
 \frac{1}{g_A^2} = 1 + \frac{2 m_N^2}{\pi g_{\pi N N}^2}
 \int_{m_\pi}^{\infty} \frac{d \nu}{\nu}\,\sqrt{\nu^2-m_\pi^2} \left
 ( \sigma_{\rm tot}^{\pi^{-}p}(\nu) - \sigma_{\rm tot}^{\pi^{+}p}(\nu) \right),
 \label{awinteg}
\end{equation}
($\nu = p\cdot q /m_N$).  Note that the $I=\half$ resonances can only
show up in the $\pi^{-} p$ scattering. We have shown -- using
unitarity -- that the corresponding resonance cross sections can at
most scale as $\order{N_c^0}$ and that in fact for the Roper-like
resonances the upper bound is saturated. Taking into account the
prefactor of the integral one easily finds that the total contribution
of the Roper-like states scales as $\order{1/N_c}$ in agreement with
our derivation presented above. In the case of the $I=\threehalf$
resonances both $\sigma_{\rm tot}^{\pi^{\pm} p}$ cross sections
contribute. The net result of the $I=\threehalf$ Roper-like resonances
is still of order $\order{1/N_c}$ and negative in order to remove the
positive $\order{1/N_c}$ contribution of their corresponding $I=\half$
partners.  For the non-resonant $\pi^{\pm}$-proton scattering,
however, it can be shown \cite{Bron 93} that the cross sections cancel
to order $\order{1/N_c}$ since the $\pi^{-}$ meson is counting the
$(N_c+1)/2$ $u$ quarks of the proton, whereas the $\pi^{+}$ is
counting the respective $N_c/2$ $d$ quarks. The total contribution of
the non-resonant $\pi$-proton scattering to the Adler-Weisberger
integral \equa{awinteg} is therefore of order
$\order{1/N_c^2}$~\cite{Bron 93}.  The contributions of the $S_{11}$
and $D_{15}$ odd-parity resonances and the other resonances whose
decay width are of the order $\order{1/N_c^2}$ appear suppressed by
the order of $\order{1/N_c^3}$ as compared to the leading and
next-to-leading effects discussed above. The predicted behavior is
qualitatively in line with the one extracted from the data on the $\pi
N$ decay widths of the $j<\sevenhalf$ resonances below 2.2~$GeV$.

\vskip 6mm
\noindent{\bf Acknowledgements}

\noindent
We thank Dr.\ W.\ Broniowski for pointing out an error in the first
version of this note.
A.W.\ would like to thank Ismail Zahed for discussions.

\newpage
\begin{table}[h]
\begin{displaymath}
\begin{array}{l l l c c c c c c }\hline
I & j & P & {\rm resonance} & g_{j I}^{(\pm)}& \eta &
    {\rm R.H.S.\ of\ (1)}
       & {\rm data }
       & \Gamma_{j I}^{(\pm)}\\ \hline
\half & \half & + & N(1440) P_{11}
                    & N_c &(-)& 1/N_c & 0.147 & 1 \\
      & \half & + & N(1710) P_{11}'
                    & N_c &(-)& 1/N_c & 0.003 & 1 \\
      & \half & - & N(1535) S_{11}
                    & N_c^{-1} &(+)& 1/ N_c^3 & 0.025 & 1/N_c^2 \\
      & \half & - & N(1650) S_{11}'
                    & N_c^{-1} &(+)& 1/ N_c^3 & 0.025 & 1/N_c^2 \\
      & \threehalf & +&  N(1720) P_{13}
                    & N_c    &(+)& 1/N_c & 0.009 & 1   \\
      & \threehalf & - & N(1520) D_{13}
                    & N_c &(-)& 1/N_c^3 & 0.060   & 1/N_c^2 \\
      & \threehalf & - & N(1700) D_{13}'
                    & N_c &(-)& 1/N_c^3 & 0.004   & 1/N_c^2 \\
      & \fivehalf  & + & N(1680) F_{15}
                   & N_c^3 &(-)& 1/N_c & 0.061 & 1  \\
      & \fivehalf  & - & N(1675) D_{15}
                   & N_c & (+)& 1/N_c^3 & 0.050 & 1/N_c^2 \\
      & \sevenhalf & - & N(2190) G_{17}
                   & N_c^3 &(-)&  1/N_c^3 & 0.016 & 1/N_c^2 \\
 \threehalf
      & \half      & + & \Delta (1910) P_{31}
                   & N_c    &(-)&  1/N_c & -0.006 & 1  \\
      & \half & - & \Delta (1620) S_{31}
                    & N_c^{-1} & (+)& 1/N_c^3 & -0.010 & 1/N_c^2 \\
      & \half & - & \Delta (1900) S_{31}'
                    & N_c^{-1} & (+)& 1/N_c^3 & -0.004 & 1/N_c^2 \\
      & \threehalf & + &\Delta (1232) P_{33}
                    & N_c^\threehalf    & (+)& 1 &-0.772   & 1/N_c^2  \\
      & \threehalf & + &\Delta (1600)P_{33}'
                    & N_c    &(+)& 1/N_c & -0.037  & 1 \\
      & \threehalf & + &\Delta (1920)P_{33}''
                    & N_c    &(+)& 1/N_c & -0.005  & 1 \\
      & \threehalf & - & \Delta (1700) D_{33}
                    & N_c &(-)& 1/N_c^3 & -0.019  & 1/N_c^2  \\
      & \fivehalf  & + & \Delta (1905) F_{35}
                   & N_c^3 &(-)& 1/N_c &-0.012  & 1  \\
      & \fivehalf  & - & \Delta (1930) D_{35}
                   & N_c &(+)& 1/N_c^3 & -0.017 & 1/N_c^2 \\
      & \sevenhalf & + & \Delta (1950) F_{37}
                   & N_c^3 &(+)& 1/N_c & -0.047 & 1  \\
                    \hline
\end{array}
\end{displaymath}
\caption[tab1]{
The $N_c$ scaling behavior of the dimensionless coupling constants
$g_{j I}^{(\pm)}$, the widths $\Gamma_{j I}^{(\pm)}$ of the
$(j^\pm,\half )$ and $(j^\pm,\threehalf )$ resonances (with $j\leq
\sevenhalf $), and their contributions to the right hand side of the
Adler-Weisberger sum rule from eq. (1).  The column denoted ``data"
contains the contributions of the single resonances to the
Adler--Weisberger sum rule as obtained in exploiting the experimental
values on the $\pi N$ partial decay widths of the resonances.  For
notations and resonance data see ref.\cite{PART 92} .}
\end{table}
\end {document}